\documentclass[prb,twocolumn,showpacs,amsmath,amssymb,
superscriptaddress,floatfix]{revtex4}

\usepackage[english]{babel}
\usepackage{graphicx}
\usepackage{times}

\begin{document}

\title{Spin-charge separation and simultaneous spin and charge Kondo effect}

\author{Rok \v{Z}itko}
\affiliation{J. Stefan Institute, Ljubljana, Slovenia}

\author{Janez \surname{Bon\v ca}}
\affiliation{Faculty of Mathematics and Physics, University of
Ljubljana, Ljubljana, Slovenia}
\affiliation{J. Stefan Institute, Ljubljana, Slovenia}

\date{\today}

\begin{abstract}
  We study the spin-charge separation in a Kondo-like model for an
  impurity with a spin and a charge (isospin) degree of freedom
  coupled to a single conduction channel (the ``spin-charge'' Kondo
  model). We show that the spin and charge Kondo effects can occur
  simultaneously at any coupling strength. In the continuum (wide-band
  or weak coupling) limit, the Kondo screening in each sector is
  independent, while at finite bandwidth and strong coupling the
  lattice effects lead to a renormalization of the effective Kondo
  exchange constants; nevertheless, universal spin and charge Kondo
  effects still occur.  We find similar behavior in the two-impurity
  Anderson model with positive and negative electron-electron
  interaction and in the two-impurity Anderson-Holstein model with a
  single phonon mode. We comment on the applicability of such models
  to describe the conductance of deformable molecules with a local
  magnetic moment.
\end{abstract}


\maketitle

\newcommand{\vc}[1]{\boldsymbol{#1}}
\newcommand{\scr}[1]{{\vec{\mathcal{#1}}}}
\newcommand{\ket}[1]{|#1\rangle}
\newcommand{\bosi}{\boldsymbol{\sigma}}
\newcommand{\dow}{\downarrow}
\newcommand{\up}{\uparrow}
\newcommand{\sutwo}{\ensuremath{\mathrm{SU}(2)}}
\newcommand{\uone}{\ensuremath{\mathrm{U}(1)}}

\section{Introduction}

Recent measurements of the transport properties of individual
molecules using scanning tunneling spectroscopy, mechanically
controllable break junctions, electromigration, and other methods have
shown that the conductance strongly depends on electron-electron
interaction as well as on the vibrational properties of
electrode-molecule-electrode junctions, i.e.  on the electron-phonon
coupling \cite{stipe1998, hahn2001, yu2004a, yu2004c, yu2005,
park2000, park2002, liang2002, pasupathy2004, wahl2005}. On the
theoretical side, the role of phonons had been studied in the
framework of the Holstein \cite{hewson1980, hewson1981} and
Anderson-Holstein models \cite{hewson2002, meyer2002, jeon2003,
cornaglia2004, lee2004a, lee2004b, cornaglia2005magneto, mravlje2005,
koch2006, hwang2006}, where the charge couples linearly to the
displacement, and various models where the center-of-mass vibrations
modulate the transparency of the tunneling barriers
\cite{cornaglia2005, balseiro2006, alhassanieh2005}.

For sufficiently strong electron-phonon coupling of the Holstein type
and for suitably tuned electrostatic potential (gate voltage), the
low-energy electron configurations may consist of the empty and doubly
occupied molecular orbital, while the singly occupied states are only
virtually excited \cite{hewson2002}. When these conditions are
fulfilled, a charge equivalent of the Kondo effect occurs
\cite{hewson2002, cornaglia2004, mravlje2005, taraphder1991}.  The
effective pseudospin degree of freedom is in this case the axial
charge \cite{jones1988} or isospin $I_z=1/2 (Q-1)$, where $Q$ is the
charge (level occupancy): empty orbital corresponds to isospin down,
$I_z=-1/2$, while the doubly occupied orbital corresponds to isospin
up, $I_z=+1/2$. The problem maps to the anisotropic charge Kondo
model, where the impurity isospin couples to the isospin (charge and
pairing) density of the conduction band via an isospin equivalent of
the antiferromagnetic exchange interaction \cite{schuttler1988,
taraphder1991, cornaglia2004}. At low temperature the isospin is
screened by pairing fluctuations in the conduction band and the ground
state is a complex many-particle Fermi liquid state which is an
isospin singlet. A $\pi/2$ phase shift occurs for low-energy
quasiparticle scattering and the molecule becomes fully conductive
when the temperature is reduced below the corresponding charge Kondo
temperature $T_K^C$. In this respect, the charge Kondo effect is
equivalent to the spin Kondo effect; the only difference are the
interchanged roles of the isospin (charge and pairing) and spin
degrees of freedom.  We also remark that the charge Kondo screening of
an isospin degree of freedom (electron {\it pairing moment}) is
fundamentally different in its origin and its properties from the
electrostatic screening of a point {\it charge} by the conduction
electrons.

The Kondo model is an effective one-dimensional quantum field theory
since the magnetic impurity is assumed to couple only to a
one-dimensional continuum of conduction electron states with $s$
symmetry about the impurity site \cite{affleck1990}. Low-dimensional
field theories have unique properties due to topological restrictions
in reduced dimensionality; for example, fermions constrained to live
on a 1D line can scatter only forwards and backwards. A notable effect
in one-dimensional systems is the separation of electron spin and
charge which had been intensively studied in Luttinger liquids
\cite{voit1995}: fundamental low-energy excitations are not charged
spin-$1/2$ Fermi-liquid quasiparticles, but rather spin-$1/2$ neutral
particles (spinons) and charged spinless particles (holons). Such
behavior has been found, for example, in one-dimensional solids such
as SrCuO$_2$ \cite{kim2006srcuo2} and ballistic wires in GaAs/AlGaAs
heterostructures \cite{auslaender2005}.

Spin-charge separation also occurs in the Kondo problem. Using bosonization
techniques, conduction band fermion fields can be described in terms of
spin-up and spin-down boson fields. These bosons correspond to the
particle-hole excitations in the conduction band and they can be recombined
to form separate spin and charge fields which are essentially independent,
but subject to a gluing condition \cite{vondelft1998, zarand2000} which is
the only remnant of the charge-$1$, spin-$1/2$ nature of physical fermion
particles. In the single-impurity spin Kondo problem, the impurity spin
couples only to the spin field, while the charge field is decoupled
\cite{blume1970, affleck1990, coleman1995}. In this sense, the spin and
charge degrees of freedom are separated. As commented in
Ref.~\onlinecite{glazman1999}, the spin-charge separation allows the
quantization of spin to persist even in the case of strong coupling of a
quantum dot to the leads when the charge is no longer quantized.  This
explains why unitary conductance can be achieved in quantum dots at
relatively high (Kondo) temperature, which is essential for experimental
observability of the Kondo effect in open quantum dots
\cite{goldhabergordon1998b}.

\begin{figure}
\includegraphics[width=8cm,clip]{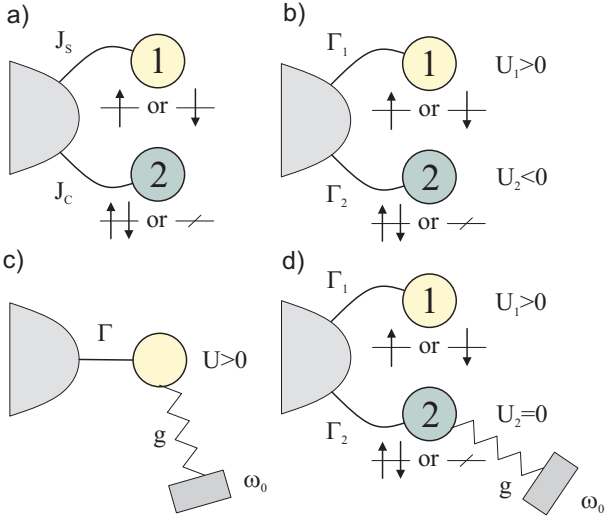}
\caption{(Color online) Schematic representations of models discussed in the paper.
a) Spin-charge Kondo model.
b) Two-impurity Anderson model with positive and negative electron-electron interaction.
c) Single-impurity Anderson-Holstein model.
d) Two-impurity Anderson-Holstein model with electron repulsion in one orbital and
electron-phonon coupling in the other.
}
\label{shemaall}
\end{figure}

An interesting situation develops when one magnetic (spin) Kondo
impurity and one charge (isospin) Kondo impurity couple to the same
conduction band, as schematically depicted in
Fig.~\ref{shemaall}a. Due to the spin-charge separation, two Kondo
screening cross-overs are expected to occur independently: the
magnetic moment on the first impurity will be screened at the spin Kondo
temperature $T_K^S \sim \exp(-1/\rho_0 J_S)$ [$\rho_0$ is the
conduction band density of states at the Fermi level, $J_S$ is the
spin Kondo antiferromagnetic exchange constant] and the pairing moment
(isospin) on the second impurity will be screened at the charge Kondo
temperature $T_K^C \sim \exp(-1/\rho_0 J_C)$, where $J_C$ is the
charge Kondo exchange constant.

The goal of this paper is to corroborate the prediction of separate
and independent spin and charge Kondo effects. We study how this
scenario breaks down in the narrow bandwidth limit (large $J_S/D$ and
$J_C/D$) when lattice effects become important. We also study the
related two-impurity Anderson model with one impurity with repulsive
electron-electron interaction, $U>0$, and one impurity with attractive
interaction, $U<0$, as we move away from the Kondo limit $|U|/\pi
\Gamma \gg 1$ by increasing the hybridization strength $\Gamma$ (see
Fig.~\ref{shemaall}b). In the Anderson model, higher $\Gamma$ implies
stronger charge fluctuation on the magnetic ($U>0$) site and stronger
spin fluctuation on the isospin ($U<0)$ site.

The paper is organized as follows. In Sec.~\ref{secII} we introduce
the spin-charge Kondo model, describe its symmetries and comment on
the possible symmetry-breaking terms. In Sec.~\ref{secIII} we present
the numerical renormalization group (NRG) results for the
thermodynamic quantities and study the degree of the spin-charge
separation.  In Sec.~\ref{secIV} we study the model in the continuum
(wide-band) limit using the non-Abelian bosonization to show how the
spin and charge degrees of freedom separate, while in Sec.~\ref{secV}
we study the opposite limit of narrow bands. This naturally leads to
the consideration of the strong-coupling fixed point in
Sec.~\ref{secVI}.  The spin-charge separation in two-impurity Anderson
models is studied in Sec.~\ref{secVII}, while in Sec.~\ref{secVIII} we
comment on the effects of the simultaneous spin and charge Kondo
effect on the transport through molecular junctions.  This is
supplemented in Sec.~\ref{secIX} by an explicit demonstration of the
charge Kondo effect in the two-impurity Anderson-Holstein model.

\section{The spin-charge Kondo model}
\label{secII}

The conventional spin Kondo model (SKM) is \cite{hewson}
\begin{equation}
H_\mathrm{SKM} = \sum_{k\mu} \epsilon_k c^\dag_{k\mu} c_{k\mu}
+J_\mathrm{S} \vc{s}(0) \cdot \vc{S},
\end{equation}
where $J_\mathrm{S}$ is the antiferromagnetic Kondo exchange constant,
\begin{equation}
\label{spin}
\vc{s}(0) = \frac{1}{2} \frac{1}{N_c^2} \sum_{kk'\mu\mu'}
c^\dag_{k\mu} \bosi_{\mu\mu'} c_{k'\mu'} = \frac{1}{2} \sum_{\mu\mu'}
f^\dag_\mu \bosi_{\mu\mu} f_{\mu'}
\end{equation}
is the conduction-band spin density at the impurity site, where
$\bosi$ are the Pauli matrices, $c^\dag_{k\mu}$ is creation operator
for conduction-band electron with momentum $k$ and spin $\mu$, $N_s$
is the number of states in the band, and $f^\dag_\mu = 1/N_c \sum_k
c_{k\mu}$ is the combination of the conduction band states which
couples to the impurity. $\vc{S}$ is the impurity spin-$1/2$ operator.
By analogy, the charge Kondo model (CKM) is \cite{taraphder1991}
\begin{equation}
H_\mathrm{CKM} = \sum_{k\mu} \epsilon_k c^\dag_{k\mu} c_{k\mu}
+ J_\mathrm{C} \vc{i}(0) \cdot \vc{I},
\end{equation}
where $J_\mathrm{C}$ is the charge Kondo exchange constant,
\begin{equation}
\label{isospin}
\vc{i}(0) = \frac{1}{2} \frac{1}{N_c^2} \sum_{kk'\alpha\alpha'}
\xi^\dag_{k\alpha} \bosi_{\alpha\alpha'} \xi_{k'\alpha'}
\end{equation}
is the conduction-band isospin density at the impurity site and
$\vc{I}$ is the impurity isospin-$1/2$ operator. Here
$\xi^\dag_{k\alpha} = \{ c^\dag_{k\uparrow}, c_{k\downarrow} \}$ is a
Nambu spinor \cite{nambu1960, anderson1958} for the conduction band
electrons, so that
\begin{equation}
\begin{split}
i_x(0) &= \frac{1}{2} \frac{1}{N_c^2} \sum_{kk'} \left(
 c^\dag_{k\uparrow} c^\dag_{k'\downarrow}
+  c_{k\downarrow} c_{k'\uparrow}
\right) \\
&= \frac{1}{2}
\left(
f^\dag_\uparrow f^\dag_\downarrow + f_\downarrow f_\uparrow
\right)
\\
i_y(0) &= \frac{1}{2} \frac{1}{N_c^2} \sum_{kk'} \left(
-i  c^\dag_{k\uparrow} c^\dag_{k'\downarrow}
+i  c_{k\downarrow} c_{k'\uparrow}
\right) \\
&= \frac{1}{2}
\left(
-i f^\dag_\uparrow f^\dag_\downarrow + i f_\downarrow f_\uparrow
\right)
\\
i_z(0) &= \frac{1}{2} \frac{1}{N_c^2} \sum_{kk'} \left(
 c^\dag_{k\uparrow} c_{k'\uparrow}
+  c^\dag_{k\downarrow} c_{k'\downarrow}
-\delta_{kk'}
\right)\\
&= \frac{1}{2}
\left(
f^\dag_\uparrow f_\uparrow + f^\dag_\downarrow f_\downarrow-1
\right).
\end{split}
\end{equation}
It should be noted that the $z$-component of the isospin density is
the electron charge density measured with respect to the half-filled
band, while $x$ and $y$-components are the electron pairing
density. Contrary to the SKM, in CKM the impurity couples only to the
charge sector of the conduction band, while the spin sector is
decoupled (see Sec.~\ref{secIV}).

We now combine the two models and introduce the spin-charge Kondo
model (SCKM) with one spin impurity and one isospin impurity, both
located at the origin:
\begin{equation}
\label{sikm}
H_\mathrm{SCKM} = \sum_{k\mu} \epsilon_k c^\dag_{k\mu} c_{k\mu}
+ J_\mathrm{S} \vc{s}(0) \cdot \vc{S}
+ J_\mathrm{C} \vc{i}(0) \cdot \vc{I}.
\end{equation}
An experimental realization of this model would consist of a single
impurity with both charge and spin degrees of freedom, such as a
deformable molecule embedded between two electrodes. In one possible
scenario, two different molecular orbitals are active. One is singly
occupied and has a magnetic moment localized, for example, on a
magnetic ion embedded in a molecule. The other orbital is as an
extended molecular orbital which is strongly coupled to a local phonon
mode.

We remark that we have assumed the isospin part of the SCKM to be
isotropic in spite of the fact that the Schrieffer-Wolff
transformation applied to the Anderson-Holstein model in general
yields an anisotropic effective exchange interaction
\cite{schuttler1988, cornaglia2004}. The exchange anisotropy in
spin-$1/2$ Kondo model is an irrelevant perturbation in the
renormalisation group sense \cite{affleck2ck1992} and we disregard
it in the first part of the paper, since it plays no role in the
context of the spin-charge separation. The anisotropy can be,
however, important in the experimental realizations of the model
since it enters the expression for the charge Kondo temperature.
We return to this point in Sec.~\ref{secIX}.

The spin-charge Kondo model is distinct from the $\sigma-\tau$ Kondo
model (also known as the compactified Kondo model) \cite{coleman1995,
coleman1995prl, bulla1997sigmatau, ye1998, bradley1999}, where a
single spin degree of freedom is coupled via $\vc{S} \cdot \vc{s}(0)$
and $\vc{S} \cdot \vc{i}(0)$ terms to a single conduction channel. In
the compactified Kondo model, one takes advantage of the spin-charge
decoupling and uses the additional isospin degrees of freedom in one
channel to mimic the spin degrees of freedom of the second channel in
the two-channel Kondo model. Our model features two different degrees
of freedom, has different symmetry and thus different properties.
Nevertheless, our results on the effects of the finite band-width
(lattice effects) on the spin-charge separation help understand why
the compactified Kondo model is not fully equivalent to the
two-channel Kondo model (see also Refs.~\onlinecite{coleman1995prl,
ye1998}).

\subsection{Symmetries and symmetry-breaking terms}
\label{secIIa}

We assume a linear dispersion relation $\epsilon_k=Dk$ for conduction
band electrons, where $2D$ is the band-width and the wave-number $k$
ranges from $-1$ to $1$.  The band is half filled (the chemical
potential is at $\mu=0$), so that we have particle-hole symmetry.  The
SCKM then has commuting $\sutwo_\mathrm{spin}$ and
$\sutwo_\mathrm{isospin}$ symmetries generated by the spin and isospin
operators
\begin{equation}
\begin{split}
s^+ &= \sum_k c^\dag_{k\uparrow} c_{k\downarrow} + S^+,
\quad s^-=(s^+)^\dag, \\
s^z &= \sum_k \frac{1}{2} \left(
c^\dag_{k\uparrow} c_{k\uparrow} - c^\dag_{k\downarrow} c_{k\downarrow}
\right) + S^Z, \\
i^+ &= \sum_k c^\dag_{k\uparrow} c^\dag_{k\downarrow} + I^+,
\quad i^-=(i^+)^\dag, \\
i^z &= \sum_k \frac{1}{2} \left(
c^\dag_{k\uparrow} c_{k\uparrow} + c^\dag_{k\downarrow} c_{k\downarrow} - 1
\right) + I^Z.
\end{split}
\end{equation}

In the presence of the magnetic field $h$ and of detuned electrostatic
potential $\delta$, we add the following perturbation terms to the
Hamiltonian:
\begin{equation}
H' = h S_z + \delta\, (Q-1) = h S_z + \delta\, (2I_z).
\end{equation}
Magnetic field breaks the $\sutwo_\mathrm{spin}$ symmetry, while
detuned electric potential breaks the $\sutwo_\mathrm{isospin}$
symmetry. Both perturbations are relevant: strong magnetic field $h >
T_K^S$ quenches the spin Kondo effect, while strong potential $\delta
> T_K^C$ quenches the charge Kondo effect.

\section{Numerical results}
\label{secIII}

We studied the model using the well-established numerical
renormalization group (NRG) technique \cite{wilson1975,
krishna1980a}. We used an implementation of the NRG where the
conservation of spin and isospin is explicitly taken into account.

We should emphasize the assumption of linear dispersion for conduction
band electrons. It is known that in one-dimensional models spin and
charge degrees of freedom truly separate only if the dispersion is
exactly linear. The separation actually extends to the region of
non-linear dispersion, however the spin and charge excitations acquire
a finite life-time \cite{haldane1981, voit1995, zacher1998,
samokhin1998}. By performing the calculations for a range of the
discretization parameters $\Lambda$ \cite{krishna1980a, campo2005}, we
have verified that the results converge rapidly to the continuous band
limit ($\Lambda \to 1$) and that the logarithmic discretization of the
conduction band does not spoil the linearity. In this work we thus
neglect the effects of non-linear conduction band dispersion, but we
note that they could in fact also be studied using suitably adapted
NRG code \cite{bulla1997}.

We computed the impurity contributions to the entropy and to the spin
and charge susceptibilities. These thermodynamic quantities are
defined as
\cite{vzporedne}
\begin{equation}
\begin{split}
S_\mathrm{imp}(T) &= \frac{\left( E-F \right)}{T}
- \frac{\left( E-F \right)_0}{T},\\
\chi_\mathrm{spin}(T) &= \frac{(g\mu_B)^2}{k_B T}
\left(
\langle S_z^2 \rangle
-
\langle S_z^2 \rangle_0
\right),\\
\chi_\mathrm{charge}(T) &= \frac{1}{k_B T}
\left(
\langle I_z^2 \rangle
-
\langle I_z^2 \rangle_0
\right),
\end{split}
\end{equation}
where the subscript $0$ refers to the situation when no impurities are
present, $E = \langle H \rangle = \mathrm{Tr} \left( H e^{-H/(k_B T)}
\right)$, $F = -k_B T \ln \mathrm{Tr} \left( e^{-H/k_B T} \right)$,
$g$ is the electronic gyromagnetic factor, $\mu_B$ the Bohr magneton
and $k_B$ the Boltzmann's constant.

\begin{figure}
\includegraphics[width=8cm,clip]{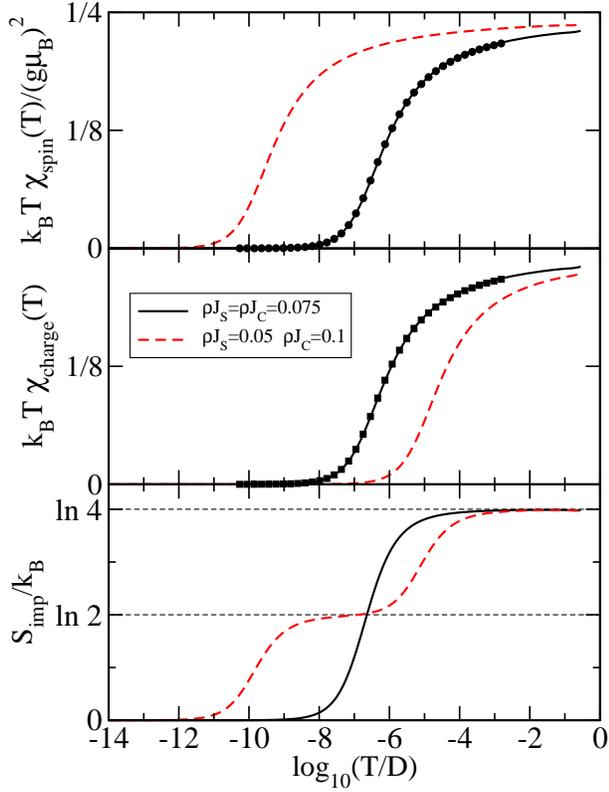}
\caption{(Color online) Impurity susceptibilities and entropy for the
spin-charge Kondo model. 
Filled symbols are a fit to the exact Bethe-Ansatz results for
susceptibility. From this fit we deduce $T_K^S=T_K^C=2.7 \times 10^{-7} D$,
in approximate agreement with $T_K=D \sqrt{\rho J}\exp(-1/\rho J)=4.4 \times
10^{-7} D$.
}
\label{fig_a}
\end{figure}

We calculated these quantities as a function of the temperature for a
number of choices of $J_S$ and $J_C$, see Fig.~\ref{fig_a}.  We used
discretization parameters $\Lambda=2,3,4$ and six different values of
parameter $z$ with the discretization scheme described in
Ref.~\onlinecite{campo2005}; the results nearly overlap for all three
$\Lambda$. At each iteration we kept states up to an energy cut-off of
at least $15 T_N$, where $T_N$ is the energy scale at the $N$-th
iteration. Even-odd effects are removed by averaging over two
consecutive iterations.

As expected, we observe Kondo screening of both spin and charge
degrees of freedom. Kondo effects in each sector appear fully
independent in the sense that the susceptibility curves
$\chi_\mathrm{spin}$ and $\chi_\mathrm{charge}$ follow the universal
$S=1/2$ Kondo forms and each is characterized by a single parameter,
the Kondo temperature $T_K^S$ or $T_K^C$, respectively. Even for
$J_S=J_C$, when the screening in each sector occurs at the same
temperature, there is no competition between the two sectors and the
curves agree perfectly with the exact Bethe-Ansatz results for the
conventional $S=1/2$ Kondo model (filled squares in Fig.~\ref{fig_a}).

\begin{figure}
\includegraphics[width=8cm,clip]{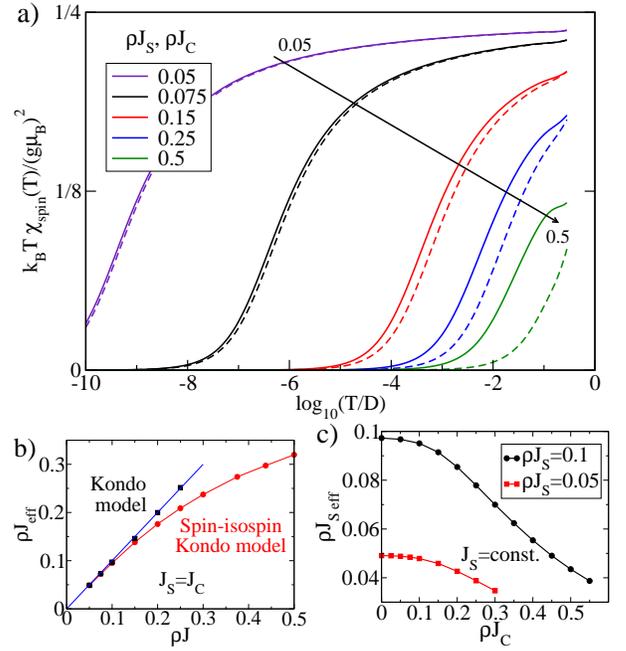}
\caption{(Color online) a) Spin susceptibilities in the SCKM
model with $J_C=J_S$ (solid lines) compared with spin susceptibilities in
the corresponding SKM with the same Kondo exchange constant $J_S$
(dashed lines). b) Effective exchange constant $J_{S\mathrm{eff}}$ as a
function of $J_S$ for the case $J_C=J_S$ (red, circles), compared
with the single-impurity results (black, squares). Blue straight line
is a fit to the single-impurity results and serves as a guide to the eye.
c) Effective exchange constant $J_{S\mathrm{eff}}$ as a
function of the charge Kondo exchange constant $J_C$ while keeping the bare 
spin exchange constant $J_S$ constant at $\rho J_S=0.1$ (black, circles) or $\rho
J_S=0.05$ (red, squares). }
\label{fig_a2}
\end{figure}

On a lattice, spin and isospin cannot simultaneously be finite at one
site \cite{coleman1995, coleman1995prl}. We therefore expect that for
large $J_C$ and $J_S$, when the Kondo temperatures are high and the
Kondo screening clouds are small, there should be some kind of mutual
disturbance. We performed a series of calculations for increasing
parameter $J=J_S=J_C$ and compared the spin susceptibility with that
of a single-impurity spin Kondo model with the same $J_S$, see
Fig.~\ref{fig_a2}a. We find that the universality is maintained: for
any $J$, the susceptibility still follows the universal Kondo curve.
The Kondo temperature is, however, reduced in the spin-charge Kondo
model and it is no longer accurately described by the familiar
expression
\begin{equation}
\label{tk}
T_K \approx D \sqrt{\rho J} e^{-\frac{1}{\rho J}}.
\end{equation}
A convenient quantity to study this renormalization effect are
effective parameters $J_{S\mathrm{eff}}$ and $J_{C\mathrm{eff}}$,
which can be obtained from respective Kondo temperatures $T^S_K$
and $T^C_K$ using the inverse of Eq.~\eqref{tk}:
\begin{equation}
\rho J_{\mathrm{eff}} = \frac{2}{W[2/(T_K/D)^2]},
\end{equation}
where $W(z)$ is the product logarithm function (the solution $x$ of
the equation $z=x \exp{x}$). In Fig.~\ref{fig_a2}b we plot the
$J_{S\mathrm{eff}}$ which corresponds to the susceptibility curves
from Fig.~\ref{fig_a2}a. For $J \ll D$ (i.e. in the wide-band limit),
there is no renormalization and $J_{\mathrm{eff}} \approx J$.  For $J
\lesssim D$ we observe a systematic reduction of the effective
exchange constant for increasing bare $J_S$. This can be understood in
terms of the ``stiffness'' of the Kondo clouds in spin and charge
sector. With increasing $J$, the screening clouds are made to occupy
smaller and smaller spatial extent around the impurity. Surprisingly,
this does not lead to a collapse of the Kondo effects, as might be
expected. Instead, the Kondo clouds spread out more than they do in
the single impurity case to compensate for the conduction electron
spin and isospin density loss due to the lattice effects.

In Fig.~\ref{fig_a2}c we show the variation of $J_{S\mathrm{eff}}$ for
a constant bare $J_S$ as the charge Kondo exchange $J_C$ is increased
from 0. The renormalization of the spin exchange constant is weaker
for smaller bare $J_S$, when the spin Kondo cloud is more spread out,
which confirms our interpretation in terms of interfering Kondo
clouds.

\section{Continuum (wide-band) limit and bosonization}
\label{secIV}

The symmetry of the problem and the separation of the spin and charge
sectors is directly exhibited in the non-Abelian bosonization approach
of the boundary conformal field theory (CFT) \cite{affleck1990,
affleck1991}. The essence of this approach is to represent the
electrons as independent bosonic fields carrying the isospin $\sutwo$
and spin $\sutwo$ degrees of freedom \cite{affleck1995, kim1997, bulla1997sigmatau}.
Introducing the left-moving field operators $\psi_{\mu}(x)$, we write
the spin-charge Kondo Hamiltonian in real space as
\begin{equation}
\label{rs}
\begin{split}
H_\mathrm{SCKM} &= \frac{i v_F}{2\pi} \sum_\mu
\int_{-\infty}^{\infty} : \psi^\dag_\mu(x)
\frac{\partial \psi_\mu(x)}{\partial x} :\, dx \\
&+ v_F \rho \left(
\lambda_S \vc{J}^S(0) \cdot \vc{S}
+ \lambda_C \vc{J}^C(0) \cdot \vc{I}
\right),
\end{split}
\end{equation}
where $v_F$ is the Fermi velocity, $\rho$ is the density of states and
$\vc{J}^S(x)$ and $\vc{J}^C(x)$ are spin and charge (isospin)
``currents'' \footnote{ In fact, $\vc{J}^S(x)$ and $\vc{J}^C(x)$ are
spin and isospin {\it densities}, but we will use the conventional CFT
nomenclature.} defined as
\begin{equation}
\begin{split}
\vc{J}^S(x) = \sum_{\mu\nu} : \psi^\dag_\mu(x)   \frac{1}{2}
\bosi_{\mu\nu} \psi_\nu(x) :, \\
\vc{J}^C(x) = \sum_{\alpha\beta} : \xi^\dag_\alpha(x) \frac{1}{2}
\bosi_{\alpha\beta} \xi_\beta(x) :
\end{split}
\end{equation}
where $\xi^\dag_\alpha(x) = \left\{ \psi^\dag_\up(x), \psi_\dow(x)
\right\}$ is the real-space Nambu spinor. Normal ordering (double
dots) has been introduced to remove divergences due to filled
electron levels below the Fermi level. Note also that in this section
the spin and charge Kondo coupling constants are $\lambda_S$ and
$\lambda_C$.

The Sugawara form \cite{goddard1986} equivalent to the Hamiltonian
Eq.~\eqref{rs} is \cite{affleck1995, kim1997}
\begin{equation}
\begin{split}
H_\mathrm{SCKM} &= \frac{\pi v_F}{l} \left[
\frac{1}{3} \sum_n : \vc{J}^S_{-n} \cdot \vc{J}^S_n :
+
\frac{1}{3} \sum_n : \vc{J}^C_{-n} \cdot \vc{J}^C_n :
\right] \\
&+\frac{\pi v_F}{l}
\left(
\lambda_S \sum_n \vc{J}^S_n \cdot \vc{S}
+\lambda_C \sum_n \vc{J}^C_n \cdot \vc{I}
\right).
\end{split}
\end{equation}
Here $\vc{J}^S_n$ and $\vc{J}^C_n$ are the Fourier modes of the
real-space currents $\vc{J}^S(x)$ and $\vc{J}^C(x)$ in a finite system
of length $2l$. They each satisfy $\sutwo_1$ Kac-Moody commutation
relations
\begin{equation}
[ J^{S/C,a}_{n}, J^{S/C,b}_{m} ] = i \epsilon_{abc} J^{S/C,c}_{n+m}
+\delta_{ab}\delta_{n+m,0} \frac{1}{2} n,
\end{equation}
where $\epsilon^{abc}$ is the antisymmetric tensor
\cite{affleck1995}. The modes from different sectors commute:
\begin{equation}
\left[ J^{S,a}_{n}, J^{C,b}_{m} \right] = 0.
\end{equation}
This commutation relation embodies the (trivial) spin-charge
separation of the free electrons ($\lambda_S=\lambda_C=0$). At a
special value $\lambda_S=1/3$, we can introduce a new current
$\scr{J}^S=\vc{J}^S + \vc{S}$ which satisfies the same Kac-Moody
commutation relations as the old currents. The spin part of the
Hamiltonian then becomes (up to a constant term)
\begin{equation}
\label{spinpart}
H^{(S)}_\mathrm{SCKM} = \frac{\pi v_F}{l} \sum_{n} \frac{1}{3}
: \scr{J}^S_{-n} \cdot \scr{J}^S_{n} :,
\end{equation}
from which the spin impurity $\vc{S}$ has disappeared (it was
``absorbed'' by the conduction band). The charge sector remains
unaffected by this change. By analogy, at a special value
$\lambda_C=1/3$ we introduce $\scr{J}^C = \vc{J}^C + \vc{I}$ and write
the charge part of the Hamiltonian as
\begin{equation}
\label{chargepart}
H^{(C)}_\mathrm{SCKM} = \frac{\pi v_F}{l} \sum_{n} \frac{1}{3}
: \scr{J}^C_{-n} \cdot \scr{J}^C_{n} :,
\end{equation}
The special values $\lambda_S=1/3$, $\lambda_C=1/3$ are identified with
the strong coupling fixed point of the problem \cite{affleck1990}.
Note that
\begin{equation}
\left[ \scr{J}^{S,a}_m, \scr{J}^{C,b}_n \right] = 0,
\end{equation}
therefore the addition of the impurities does not break the
spin-charge separation.

Even though the spin Kondo effect occurs in the spin sector, without
involving the charge sector, and the charge Kondo effect occurs in the
charge sector, without involving the spin sector, the spin and charge
degrees of freedom are not entirely decoupled; they are constrained by
the gluing condition. The gluing condition declares which combinations
of quantum numbers are allowed taking into account the charge-1
spin-1/2 nature of physical particles -- electrons. In the present
context the gluing condition depends on the boundary conditions
(b. c.)  imposed on the field $\psi(x)$.

We first consider the case of anti-periodic b. c., $\psi(l) =
-\psi(-l)$. We can obtain half-integer spin only with an odd number of
electrons (i.e. for half-integer isospin)
\cite{affleck1990}. Therefore $2I^z$ and $2S^z$ must have the same
parity; this is the gluing condition. There are thus two Kac-Moody
conformal towers with highest-weight states having $(I,S)=(1/2,1/2)$
and $(I,S)=(0,0)$, respectively.

For periodic b.c. $\psi(l)=\psi(-l)$, and keeping in mind that the
$z$-component of the isospin is defined with respect to half filling,
we obtain half-integer spin for integer isospin and integer spin for
half-integer isospin; $2I^z$ and $2S^z$ must then have different
parity. There are two conformal towers, $(I,S)=(1/2,0)$ and
$(I,S)=(0,1/2)$. Note that changing the b. c. from periodic to
anti-periodic (or vice versa) amounts to imposing a phase shift of
$\pi/2$ on the wave function \cite{affleck1991}.

In the single impurity Kondo model, the finite-size spectrum of the
strong coupling fixed point is obtained by a fusion in the spin sector
\cite{affleck1990, affleck1991}. This means that the isospin sector
remains intact, while the spin quantum number changes as $S \to
1/2-S$. As a consequence $(1/2,1/2) \to (1/2,0)$ and $(0,0) \to
(0,1/2)$, i.e. the gluing conditions change from those for the
anti-periodic b. c. to those for periodic b. c., and vice versa, which
is equivalent to the $\pi/2$ phase shift.

We now generalize this fusion rule to the case of both (spin and
isospin) impurities. The absorption of the spin impurity will not
affect the isospin sector ($I \to I$), however $S \to
1/2-S$. Similarly, the absorption of the isospin impurity does not
affect the spin sector $S \to S$, while $I \to 1/2-I$. The fusion rule
is thus $S \to 1/2-S$ and $I \to 1/2-I$. Therefore $(0,0) \to
(1/2,1/2)$ and $(1/2,1/2) \to (0,0)$ for periodic b.c. and $(1/2,0)
\to (0,1/2)$ and $(0,1/2) \to (1/2,0)$ for anti-periodic b.c.: the
boundary conditions remain the same, only the conformal towers are
permuted. There is no phase shift in this case. Alternatively, we can
consider the fusion in each sector to give a $\pi/2$ phase shift. This
gives a total phase shift of $\pi$ which, however, is equivalent to
zero phase shift since phase shifts are defined modulo $\pi$. The
finite-size spectrum is that of free fermions, therefore the fixed
point corresponds to a Fermi liquid.

\section{Zero-bandwidth and narrow-band limits}
\label{secV}

For very large $J_C$ and $J_S$, SCKM reduces to the zero-bandwidth
limit where the band is effectively described as a single orbital. In
this three-site problem, the ground state for $J_S>J_C$ is a spin
singlet + a free isospin, while for $J_C>J_S$ the ground state is an
isospin singlet + a free spin. Spin and isospin are in direct
competition since a single band ``orbital'' can either behave as a
spin degree of freedom or as an isospin degree of freedom, but not
both at the same time. Only one of the two possible Kondo ground
states can be realized.

At finite bandwidth additional degrees of freedom become available and
the spin and charge Kondo effects can occur simultaneously.  In the
next approximation, we therefore take two lattice sites to mimic the
conduction band. This ``narrow-band limit'' is the minimal model for a
band with independent spin and isospin degrees of freedom. We use
$H_\mathrm{band}=-t \sum_\mu c^\dag_{1\mu} c_{0\mu} +
\text{H.c.}$, where $t \ll J_C, J_S$. In this approximation the ground
state is spin-singlet isospin-singlet for any values of $J_C$ and
$J_S$, as expected.  The ground state energy is
\begin{equation}
E_\mathrm{GS} = -\frac{3}{8} \left( J_C + J_S + \sqrt{(J_C-J_S)^2 +
\left(\frac{8}{3}\right)^2 t^2} \right).
\end{equation}
which, for large $J_S/t, J_C/t$, equals
\begin{equation}
E_\mathrm{GS} \sim
\begin{cases}
- \frac{3}{4} \max\{ J_C,J_S \} - \frac{4}{3} \frac{t^2}{|J_C-J_S|}
& \text{if}\, |J_C-J_S| \gg t, \\
- \frac{3}{4} J - t & \text{if}\, |J_C-J_S| \ll t. \\
\end{cases}
\end{equation}
In the second line $J\sim J_C \sim J_S$. The fixed point Hamiltonian
consists of the impurity sites strongly coupled to the first two sites
of the Wilson chain, which thereby become decoupled from the remainder
of the chain.

\section{Fixed points}
\label{secVI}

\begin{figure}
\includegraphics[width=6cm,clip]{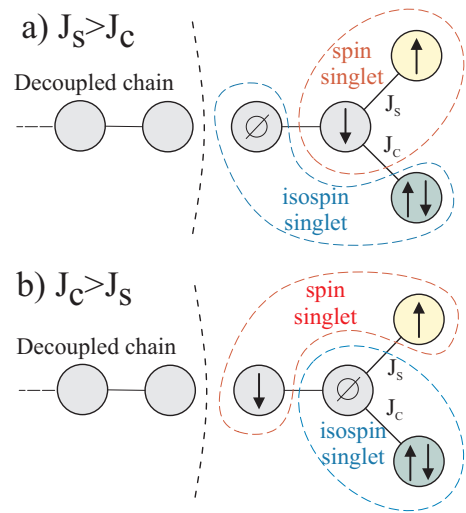}
\caption{(Color online) Schematic representation of the strong
  coupling fixed point Hamiltonian of the spin-charge Kondo model for
  a) $J_S > J_C$ and b) $J_C>J_S$. For $J_C=J_S$, the fixed point
  corresponds to the symmetric combination of a) and b). }
\label{fixedpoint}
\end{figure}

In the single-impurity Kondo model with one spin (isospin) impurity,
the strong coupling fixed point corresponds to the impurity site and
the first site of the Wilson chain being tightly bound into a spin
(isospin) singlet, which results in the decoupling of the first site
of the Wilson chain from the rest \cite{wilson1975, krishna1980a}. As
a result, the low-energy electrons experience a $\pi/2$ scattering
phase shift. As shown in the previous section, in the strong coupling
fixed point of the SCKM two sites are removed from the chain (see
Fig.~\ref{fixedpoint}). This corresponds to a zero phase shift for
quasiparticles, in agreement with the CFT analysis.

The $\sutwo_\mathrm{spin} \times \sutwo_\mathrm{isospin} \simeq
\mathrm{O}(4)$ symmetry with zero phase shift of the SCKM model should
be contrasted with the $\mathrm{O}(3) \times \mathrm{O}(1)$ symmetry
of the $\sigma-\tau$ model at its NFL fixed point or the
$\mathrm{O}(4)$ symmetry with $\pi/2$ phase shift at its FL fixed
point when an anisotropy between spin and isospin sectors exists
\cite{ye1998, bulla1997sigmatau}.

\section{Spin-charge separation in the two-impurity Anderson model}
\label{secVII}

The Kondo model is an effective low-temperature theory of a magnetic
impurity. If the physical reality is modelled more accurately, the
impurity is described using the single impurity Anderson model (SIAM)
\cite{anderson1961} and the Kondo model arises only as a
low-temperature effective theory of SIAM after the charge-fluctuations
are projected out using the Schrieffer-Wolff transformation
\cite{schrieffer1966}. In SIAM, charge fluctuations persist at low
temperatures and, in fact, they must be present since they provide the
mechanism by which the impurity spin can flip.  Nevertheless, in the
strong Kondo regime, $U/\pi \Gamma \gg 1$ (where $U$ is the electron
repulsion and $\Gamma$ the hybridization strength), the Anderson
impurity still couples predominantly to the spin sector of the
conduction band and only weakly to the charge sector. In fact, this
problem has to be considered from the renormalization group point of
view. At high energy (temperature), when the system is in the
free-orbital regime, the impurity indeed couples to both spin and
charge sectors. At low energy, when the system is in the local moment
regime, the coupling to the charge sector is frozen out.  This implies
that the spin and charge separate only for low energy electrons, while
the two degrees of freedom are ``entangled'' for high energy
electrons. Due to the energy-scale separation in quantum impurity
models \cite{wilson1975}, the lack of the spin-charge separation at
high energies does not preclude the spin-charge separation at low
energies (i.e. at low temperatures).  Alternatively, this can be
phrased in terms of the separation of time scales \cite{vzporedne}:
the duration of fluctuations (spin flips) is $\tau_U \sim \hbar/U$
while the ``magnetic'' time scale (roughly equivalent to the mean time
between successive spin flips) is $\tau_K \sim \hbar/T_K$. In the
Kondo limit, $T_K \ll U$ and therefore $\tau_K \gg \tau_U$.

In the Anderson model with negative $U$ the spin and charge sectors
are interchanged \cite{taraphder1991}. For temperatures below $|U|$, a
pairing moment develops in place of the magnetic moment. The impurity
state can be flipped from zero-occupancy (isospin down) to
double-occupancy (isospin up) by coupling to the conduction band. The
single-occupancy states can be projected out using a suitably
generalized Schrieffer-Wolff transformation and the effective model is
the charge (isospin) Kondo model \cite{taraphder1991}.

Generalizing the spin-charge Kondo model, we now study the
two-impurity Anderson model (2IAM) described by the Hamiltonian
\begin{equation}
\begin{split}
H &= \sum_{k\sigma} \epsilon_k c^\dag_{k\sigma} c_{k\sigma}
+ \frac{U_1}{2} (n_1-1)^2 + \delta_1 (n_1-1) \\
&+ \frac{U_2}{2} (n_2-1)^2 + \delta_2 (n_2-1) \\
&+ \sum_{k\sigma} V_k^1
\left( d^\dag_{1\sigma} c_{k\sigma} + \text{H.c.} \right)
+ \sum_{k\sigma} V_k^2
\left( d^\dag_{2\sigma} c_{k\sigma} + \text{H.c.} \right).
\end{split}
\end{equation}
Here $n_i=\sum_\sigma d^\dag_{i\sigma} d_{i\sigma}$ is the electron
occupancy of impurity $i$, $U_1>0$ is the electron-electron (e-e)
repulsion on impurity 1, $U_2<0$ is the e-e attraction on impurity 2.
We assume a constant hybridization strength $\Gamma=\pi \rho_0
|V_{k_F}|^2$; this permits comparison with the Kondo model which, in a
similar manner, has a constant density of state $\rho_0$ and a
constant exchange constant $J_K$ (this corresponds to approximation
$V_k=V_{k_F}$). Parameters $\delta_i = \epsilon_i + U_i/2$ measure the
departure from the particle-hole (p-h) symmetric point at
$\delta_1=\delta_2=0$.  The model is schematically depicted in
Fig.~\ref{shemaall}b. In the p-h symmetric point, 2IAM has
$\sutwo_\mathrm{spin} \times \sutwo_\mathrm{isospin}$ symmetry like
SCKM.

\begin{figure}
\includegraphics[width=8cm,clip]{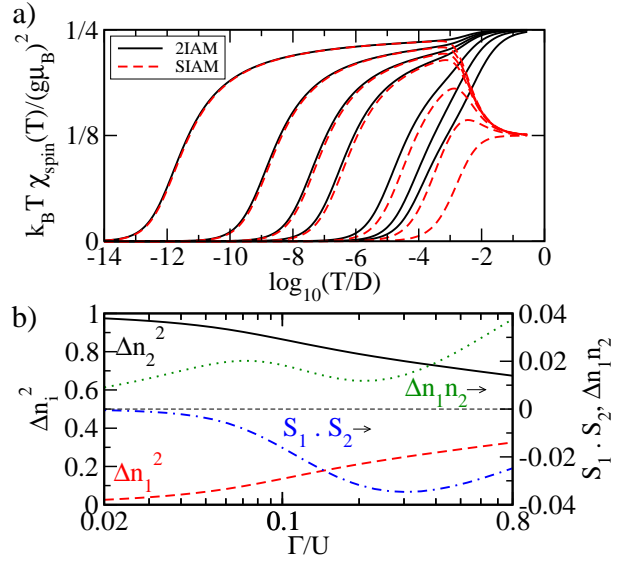}
\caption{(Color online) a) Spin susceptibility for 2IAM (full lines).
We choose $U_1=-U_2=U$ and $\Gamma_1=\Gamma_2=\Gamma$. For comparison we
plot results of the SIAM with the same $\Gamma$ and $U$ (dashed lines). Parameters are
$U/D=0.01$ and $\Gamma/U=0.02, 0.03, 0.04, 0.05, 0.1, 0.2$ and $0.5$ (left
to right).  b) Charge fluctuations and correlations in 2IAM for increasing
hybridization strength $\Gamma$.
}
\label{fig_b}
\end{figure}

In Fig.~\ref{fig_b} we show the spin susceptibility and expectation
values of the local charge fluctuations $\Delta n_i^2 = (n_i- \langle
n_i \rangle)^2$, inter-impurity charge fluctuations $\Delta n_1 n_2=
(n_1-\langle n_1 \rangle)(n_2-\langle n_2 \rangle)$, and the
inter-impurity spin correlations $\vc{S}_1 \cdot \vc{S}_2$ as a
function of the hybridization strength $\Gamma$.

For small $\Gamma$, i.e. for small equivalent Kondo exchange couplings
\begin{equation}
J_S = \frac{8\Gamma_1}{\pi U_1},\quad J_C = \frac{8\Gamma_2}{\pi|U_2|},
\end{equation}
the first impurity behaves at low temperatures as the SIAM with the
same $U_1, \Gamma_1$, see Fig.~\ref{fig_b}a. The good agreement at low
temperatures demonstrates that spin and charge degrees of freedom in
the 2IAM effectively separate in spite of charge fluctuation on the
impurity 1 and spin fluctuations on the impurity 2, as discussed above
for the case of a single Anderson impurity.  At temperatures above the
moment formation temperature $T^* \approx 1/5 U$ \cite{krishna1980a},
the magnetic susceptibility is larger than that of the SIAM model by a
factor of 2 as both impurities are then in the free orbital regime
with equally probable configurations $\ket{0}, \ket{\uparrow},
\ket{\downarrow}, \ket{\uparrow\downarrow}$.

For increasing $\Gamma/U$, the renormalization effect due to the
presence of impurity 2 becomes noticeable and the Kondo temperature is
lower than in the corresponding SIAM model. This can again be
explained in terms of Kondo cloud ``stiffness'', see
Sec.~\ref{secIII}. For very large $\Gamma/U$ we observe an interesting
saturation effect. While for $U/\pi\Gamma < 1$ the Kondo effect in
SIAM collapses, charge fluctuations rise to the maximum value of
$\Delta n^2=0.5$ and the susceptibility drops to zero at $T \sim
\Gamma$ \cite{krishna1980a}, in 2IAM the charge fluctuations on
impurity 1 are constrained due to the presence of impurity 2 (see the
slow rise of the $\Delta n_1^2$ curve in Fig.~\ref{fig_b}b) and,
similarly, spin fluctuations on impurity 2 are constrained due to the
presence of impurity 1 (see $\Delta n_2^2$ curve). The temperature
dependence of the susceptibility is in this case approximately
$T\chi(T) \sim \log{T}$ in a temperature interval of four orders of
magnitude, with the low-temperature tail still having the form of the
universal Kondo curve.

\begin{figure}
\includegraphics[width=8cm,clip]{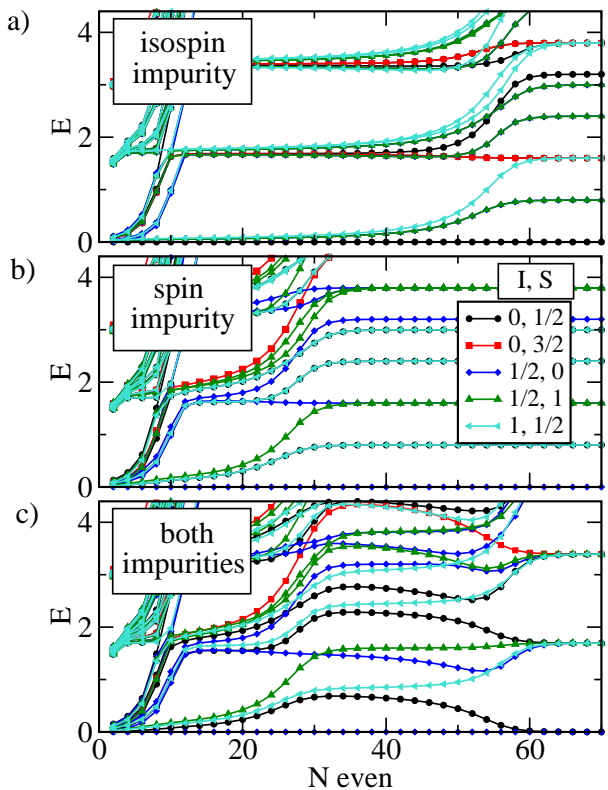}
\caption{(Color online) Numerical renormalization group eigenvalue flows for
the case when a) isospin impurity only, b) spin impurity only, c) both
impurities are present. $U_1=U=0.01$, $\Gamma,{1,2}/U=0.05$ (unless turned
to zero in a and b), $U_2=-3U$. The levels are labeled by the total isospin
and spin quantum numbers, $(I,S)$.
}
\label{fig_d3}
\end{figure}

In Fig.~\ref{fig_d3} we show NRG eigenvalue flows for: a) single $U<0$
impurity, b) single $U>0$ impurity and c) the 2IAM with both
impurities. It clearly shows the difference between SIAM and 2IAM: for
negative-$U$ SIAM the ground state is an isospin singlet and the
system is a Fermi liquid with a $\delta_\mathrm{qp}=\pi/2$ phase shift; for
positive-$U$ SIAM the ground state is a spin singlet and the system is
a Fermi liquid with a $\delta_\mathrm{qp}=\pi/2$ phase shift; finally, in the 2IAM
the ground state is a spin-singlet isospin-singlet (for odd $N$; for
even $N$ that we show, the finite-size ground state consists of
degenerate $I=1/2,S=0$, $I=0,S=1/2$ states) and the system is a Fermi
liquid with zero phase shift as in the SCKM model, see
Sec.~\ref{secVI}.

\begin{figure}
\includegraphics[width=8cm,clip]{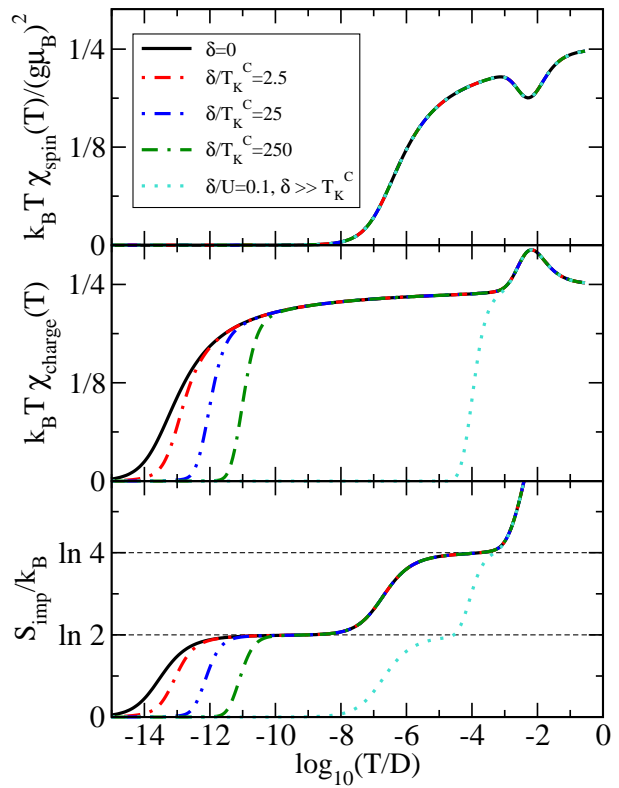}
\caption{(Color online) Impurity susceptibilities and entropy for the
two-impurity Anderson model. Impurity $1$ has $U_1=U=0.01$, $\Gamma/U=0.05$.
Impurity $2$ has $U_2=-3U=-0.03$ and equal hybridization $\Gamma$. $T_K^S/D
\sim 1.1\ 10^{-6}$, $T_K^C/D \sim 2.9\ 10^{-13}$. Note that the magnetic
susceptibility curves nearly overlap for all values of $\delta$.
}
\label{fig_d2}
\end{figure}

The departure from the p-h symmetric point is a marginal perturbation
for the positive-$U$ impurity and, as long as $|\delta_1| \lesssim
U/2$, the spin Kondo effect persists \cite{krishna1980a}. The p-h
symmetry breaking is, however, a relevant perturbation for the
negative-$U$ impurity, where it plays the same role as a magnetic
field for the positive-$U$ case (see Sec.~\ref{secIIa}). If
$|\delta_2| > T_K^C$, the charge Kondo effect is washed out. This is
illustrated in Fig.~\ref{fig_d2}, where we plot the thermodynamic
quantities for a range of $\delta=\delta_1=\delta_2$. It should be
noticed that the spin susceptibility is hardly affected; the curves
nearly overlap. The effect of a magnetic field is analogous: it washes
out the spin Kondo effect, but it is marginal for the charge Kondo
effect.

In Fig.~\ref{fig_d2} we observe some high temperature features in the
thermodynamic properties at $T \sim U_1, U_2$, e.g., a dip in the spin
susceptibility and a peak in the charge susceptibility. This is a
common feature of the 2IAM models with $U_1 \neq -U_2$. In a single
impurity model with $U_1>0$, the spin susceptibility increases at the
local moment formation temperature $T_1^*=1/5U_1$ from its free
orbital value of $1/8$ to the local moment value of $1/4$, while the
charge susceptibility sharply drops to zero as the charge fluctuations
are frozen out. In a single impurity model with $U_2<0$ the same
scenario occurs at $T_2^*=1/5 |U_2|$ with the spin and isospin sectors
exchanged. Therefore, in 2IAM with $U_1 < |U_2|$, as in this case, we
observe a dip in the spin susceptibility and a peak in the charge
susceptibility, since the pairing moment begins to form at a higher
temperature than the magnetic moment. For $U_1 > |U_2|$, there is a
peak in the spin susceptibility and a dip in the charge
susceptibility.  Finally, for $U_1 = -U_2$, magnetic and pairing
moments are formed at the same temperature and the variations in
$\chi_\mathrm{spin}$ and $\chi_\mathrm{charge}$ are hardly visible.

\section{Spectral densities and conductance}
\label{secVIII}

Using NRG one can calculate the frequency dependence of the
single-particle spectral densities \cite{costi1993, costi1994,
hofstetter2000, costi2001}, which determine the conductance through a
deformable molecule with spin and charge degrees of freedom. We make a
simplifying assumption that the molecule described by 2IAM is
symmetrically embedded between two electrodes, so that the conduction
channel in the model corresponds to the symmetric combination of
electrons from both electrodes \cite{glazman1988}. The coupling is
then proportionate and we may use the Meir-Wingreen formula for the
conductance \cite{meir1992}
\begin{equation}
\label{mw}
G=G_0 \int_{-\infty}^{\infty} \frac{\partial f}{\partial \omega}
\mathrm{Im}[\mathrm{Tr}\{ \boldsymbol{\Gamma}
{\bf{G}}^r \} ] d{\omega},
\end{equation}
where $G_0=2e^2/h$ is the conductance quantum, $f$ the Fermi-Dirac
distribution function, $\boldsymbol{\Gamma}$ the hybridization matrix
and ${\bf{G}}^r$ the retarded Green's function matrix. The components
of the hybridization matrix are
\begin{equation}
\Gamma_{i,j} = \pi \rho(\omega) V_i(\omega) V^*_j(\omega),
\end{equation}
where $V_n$ is the hopping amplitude from the impurity orbital $n$ to
the conduction band. In a simplified model we assume a constant
density of states $\rho$ and an energy-independent hybridization
strength $\Gamma=\pi \rho_0 |V|^2$ which is the same for both
orbitals. All components of the hybridization matrix are then the
same: $\Gamma_{ij} = \Gamma$. Equation~\eqref{mw} simplifies to
\begin{equation}
G=G_0 \int_{-\infty}^{\infty}
\left( - \frac{\partial f}{\partial \omega} \right)
\pi \Gamma \sum_{ij} \mathrm{Im} \left(-\frac{1}{\pi}
G^r_{ij}\right) d{\omega}.
\end{equation}
The quantity in the parenthesis is related to the spectral density
matrix for both orbitals, $A_{ij}=-1/(2\pi) \mathrm{Im} (G^r_{ij}+G^r_{ji})$. It can be
computed using standard NRG techniques from matrix elements of the
creation operators using the following spectral decompositions:
\begin{equation}
\begin{split}
A_{i,j}(\omega>0) &= \frac{1}{Z} \sum_{m,n_0} \mathrm{Re}
\left[
\left( \langle m | d^\dag_i | n_0 \rangle \right)^*
\langle m | d_j^\dag | n_0 \rangle
\right] \\
& \quad \times \delta(\omega-E_m), \\
A_{i,j}(\omega<0) &= \frac{1}{Z} \sum_{m_0,n} \mathrm{Re}
\left[
\left( \langle m_0 | d^\dag_i | n \rangle \right)^*
\langle m_0 | d_j^\dag | n \rangle
\right] \\
& \quad \times \delta(\omega+E_n),
\end{split}
\end{equation}
where $Z$ is the spectral sum $Z=\mathrm{Tr}(e^{-\beta H})$, indices
$m_0,n_0$ with subscript $0$ run over (eventually degenerate) ground
states and indices $m,n$ without a subscript over all states. Delta
functions need to be appropriately broadened \cite{bulla2001}. Note
that there is a sum rule
\begin{equation}
\int_{-\infty}^{\infty} A_{ij}(\omega) d\omega = \delta_{ij},
\end{equation}
which follows from the fermionic anti-commutation relation
$a^\dag_{i\mu} a_{j\mu} + a_{j\mu} a^\dag_{i\mu} = \delta_{ij}$.

We are particularly interested in the symmetrized and normalized spectral
density function $g(\omega)$ defined by
\begin{equation}
g(\omega)=\pi \Gamma \sum_{ij} A_{ij}(\omega)
\end{equation}
where $i,j=1,2$. This quantity appears in the final expression for the
conductance:
\begin{equation}
G = G_0 \int_{-\infty}^{\infty} \left( -\frac{\partial f(\omega)}
{\partial \omega} \right) g(\omega) d\omega.
\end{equation}
In a simple approximation, the temperature dependence of the
conductance through the quantum dots can be deduced from the energy
dependence of the function $g(\omega)$.

\begin{figure}
\includegraphics[width=8cm,clip]{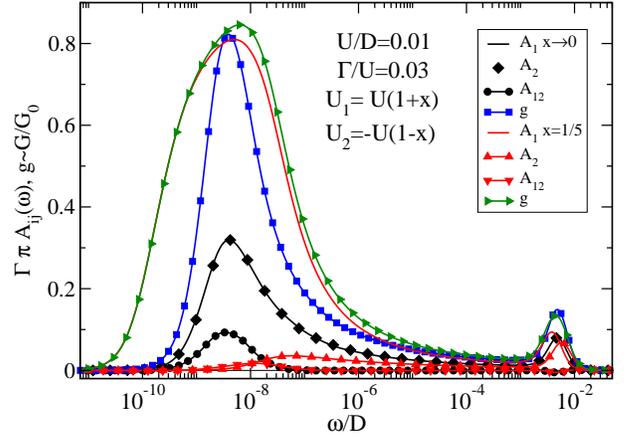}
\caption{(Color online) Diagonal ($A_{1}$, $A_{2}$) and out-of-diagonal
spectral density ($A_{12}$) of the two impurity Anderson model for different
asymmetry parameters $x$. We plot only the positive frequencies. Due to the
particle-hole symmetry, we have $A_i(\omega)=A_i(-\omega)$ and
$A_{12}(\omega)=-A_{21}(-\omega)$.
}
\label{fig_d}
\end{figure}

We computed the spectral densities $A_i(\omega)=A_{ii}(\omega)$ and
the out-of-diagonal spectral density $A_{12}(\omega)$ for a constant
$\Gamma$ and a range of $U_1,U_2$, defined by $U_1=U(1+x)$ and
$U_2=-U(1-x)$ with constant $U$, see Fig.~\ref{fig_d}. For $x=0$, the
Kondo temperatures are the same.

In the generic case $x \neq 0$ (assuming $x > 0$, so that $T_K^S >
T_K^C$), the spectral density of the first impurity increases for
$\omega \lesssim T_K^S$, but then it drops to zero for $\omega
\lesssim T_K^C$, see the case of $x=1/5$. We thus obtain a peak
centered at $\omega \sim \sqrt{T_K^S T_K^C}$. This is at first
surprising, since spin Kondo effect is usually associated with an
Abrikosov-Suhl resonance at zero frequency in the spectral density
and, due to the spin-charge decoupling, one would expect no influence
of the charge Kondo effect induced by the second impurity. Indeed, at
low temperatures the thermodynamic functions related to spin are
universal functions of $T_K^S$ only; for example, $\chi_\mathrm{spin}$
in no way depends on the charge Kondo temperature $T_K^C$. There is
clearly no such universality in the dynamic quantities. Spectral
density is related to the injection of a physical fermion in the
system; the fermion carries both spin and isospin degrees of freedom
and is therefore sensitive to both sectors.

For relatively large $x=1/5$, the spectral density on the second
impurity, $A_{2}(\omega)$, has a small bulge at $T_K^S$ and then it
drops to zero. We emphasize that no Abrikosov-Suhl resonance appears
in $A_{2}(\omega)$ below $T_K^C$ as might be expected. For $x$ tending
towards 0, the bulge in $A_2(\omega)$ develops into a peak of the same
form as the one found in $A_1(\omega)$. The out-of-diagonal spectral
density $A_{12}(\omega)$ also builds up a small peak at $\omega \sim
T_K$ as $x \to 0$ and goes to zero at low frequencies for all
$x$. These results indicate that the temperature dependence of
conductance is expected to be non-monotonic: it begins to increase at
the higher of the two Kondo temperatures, but then it decreases when
the lower Kondo temperatures is reached. The system does not conduct
at zero temperature, $g(\omega \to 0)=0$. This results is consistent
with the zero temperature conduction of Fermi liquids as determined by
the quasiparticle phase shift $\delta_\mathrm{qp}=0$:
\begin{equation}
G(T=0)=G_0 \sin^2 \delta_\mathrm{qp}=0.
\end{equation}

\begin{figure*}
\includegraphics[width=16cm,clip]{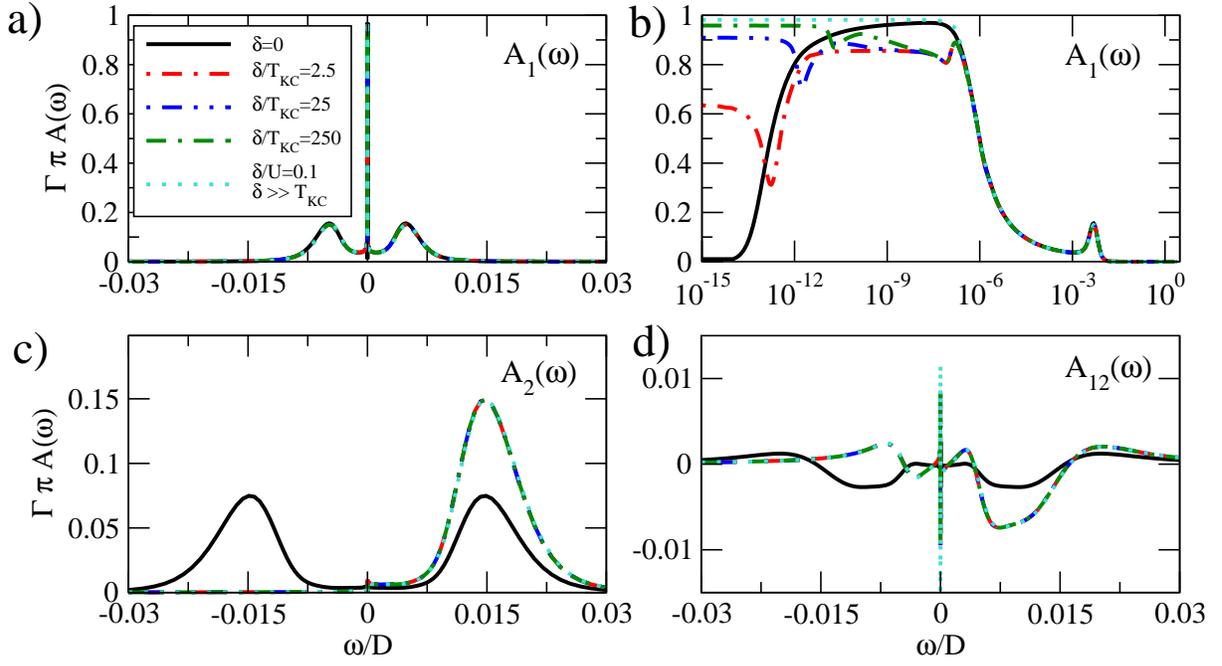}
\caption{(Color online) Diagonal ($A_{1}$, $A_{2}$) and out-of-diagonal
spectral densities ($A_{12}$) of the two impurity Anderson model.
Same parameters as in Fig.~\ref{fig_d2}.
}
\label{fig_d2spec}
\end{figure*}

The effect of the gate voltage $\delta$ is shown in
Fig.~\ref{fig_d2spec} for the case of widely separated spin and charge
Kondo temperatures, $T_K^C \ll T_K^S$ (see also Fig.~\ref{fig_d2}
where the thermodynamic properties are shown for the same parameters).
For $\delta > T_K^C$, the isospin fluctuations on the second orbital
are rapidly quenched and the isospin degree of freedom is fully
polarized [see $A_{2}(\omega)$]. The low-frequency peak in the
spectral density on the first orbital is restored at the same time. It
should be noted that the out-of-diagonal spectral density
$A_{12}(\omega)$ is small for all $\delta$. With increasing $\delta$,
the conductance thus rises from zero toward the unitary limit
\begin{equation}
G(\delta \gg T_K^C) \approx G_0.
\end{equation}
We consider the significance of such behavior in the conclusion.

\section{Two-impurity Anderson-Holstein model}
\label{secIX}

In previous sections, we performed calculations using the 2IAM with a
negative $U_2$ parameter that we considered as a given constant. In
this section we first study how an effective negative $U_2$ emerges in
the presence of the charge-phonon coupling (Anderson-Holstein model)
and then extend this model to the two impurity case with one Anderson
impurity (spin degree of freedom) and one Holstein impurity (isospin
degree of freedom).

The Anderson-Holstein model is an extension of the Holstein model in that it
considers electrons with spin and (optionally) a finite on-site
electron-electron repulsion in the impurity \cite{hewson2002}, see
Fig.~\ref{shemaall}c. The Hamiltonian is
\begin{equation}
\begin{split}
H &= \sum_{k\sigma} \epsilon_k c^\dag_{k\sigma} c_{k\sigma}
+ \sum_{k\sigma} V_k \left(
d^\dag_\sigma c_{k\sigma} + \text{H.c.}
\right) \\
& + U (n-1)^2 + g (a^\dag+a) (n-1) + \omega_0 a^\dag a
\end{split}
\end{equation}
where $d$ is the electron annihilation operator on the impurity and
$a$ is the annihilation operator for a phonon of frequency $\omega_0$
which couples to the charge $n=\sum_{\sigma} d^\dag_\sigma d_\sigma$
with electron-phonon (e-ph) coupling strength $g$. The phonon does not
break the particle-hole symmetry, however the full isospin
$\sutwo_\mathrm{isospin}$ symmetry is reduced to
$\uone_\mathrm{charge}$ conservation of charge symmetry.

In general, the phonon renormalizes both the e-e interaction $U$,
which becomes frequency dependent, and the hybridization strength
$\Gamma$. In the anti-adiabatic limit $\omega_0 \gg U$, the effective
e-e interaction again becomes instantaneous and the Anderson-Holstein
model maps to the usual single impurity Anderson model (SIAM) with
\cite{hewson2002}
\begin{equation}
\label{Ueff}
U_\mathrm{eff} = U-2g^2/\omega_0.
\end{equation}
For large enough $g$, $U_\mathrm{eff}$ becomes negative and the charge
Kondo effect is expected. In this limit ($\omega_0 \gg U$), the
effective exchange interaction for the isospin degrees of freedom is
isotropic \cite{hwang2006} and the $\sutwo_\mathrm{isospin}$ symmetry
is largely restored at energies much lower than $\omega_0$.

\begin{figure}
\includegraphics[width=8cm,clip]{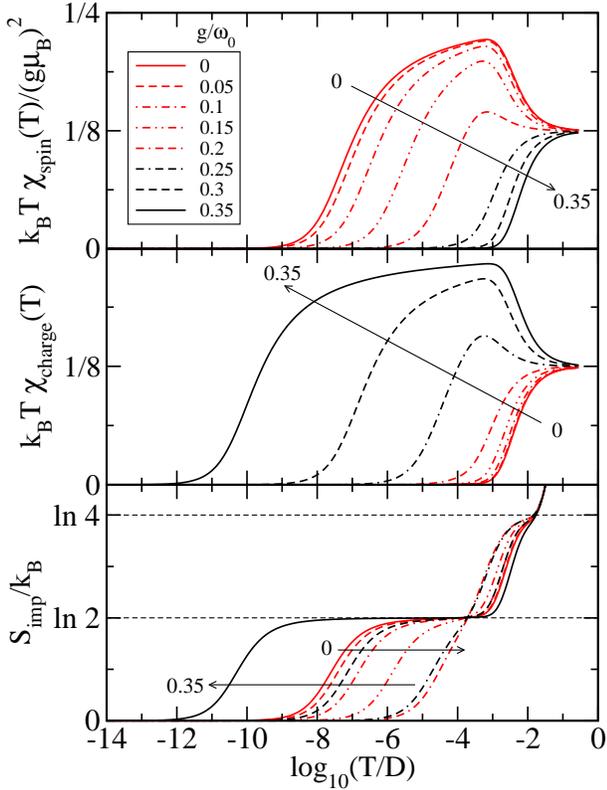}
\caption{(Color online) Impurity susceptibilities and entropy for the
single-impurity Anderson-Holstein model for increasing e-ph interaction
strength $g$. $U/D=0.01$, $\Gamma/U=0.04$, $\omega_0/D=0.1$.
}
\label{fig_e}
\end{figure}

The reversal of roles of spin and charge fluctuations in the
Anderson-Holstein model was first deduced from dynamic spin and charge
susceptibilities at zero temperature in Ref.~\onlinecite{hewson2002}.
In Fig.~\ref{fig_e} we demonstrate this reversal in the temperature
dependent thermodynamic spin and charge susceptibilities where the
disappearance of the magnetic moment and emergence of the pairing
moment is explicit.

When the Coulomb repulsion in an orbital is small, $U \to 0$, but the
coupling to a phonon mode is substantial, we have $U_\mathrm{eff} =
-2g^2/\omega_0$. This holds {\it exactly} in this case (i.e. the
condition $\omega_0 \gg U$ is trivially met).
If, however, $U$ and $\omega_0$ in the same orbital are of the same
magnitude, or $U \gg \omega_0$, the appropriate mapping is to the
anisotropic Kondo model \cite{schuttler1988, cornaglia2004}. In such
systems the charge Kondo temperature is generally strongly attenuated,
therefore in this work we do not discuss the case $U \gtrsim \omega_0$
further.

\begin{figure}
\includegraphics[width=8cm,clip]{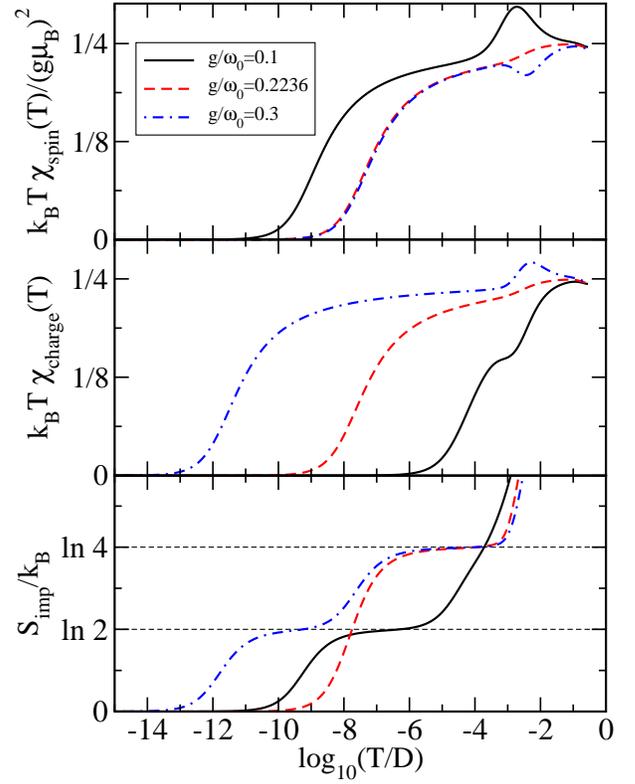}
\caption{(Color online) Impurity susceptibilities and entropy for the
two-impurity model with one Holstein impurity with $\omega_0/D=0.1$ and varying
e-ph interaction strength $g$, and one Anderson impurity with $U/D=0.01$.
Both impurities are coupled with $\Gamma/D=0.0004$ to the conduction band.
}
\label{fig_f3}
\end{figure}

As an illustration, we now consider the case of two active orbitals,
the first with $U_1>0$ and the second with $U_2=0$ and a phonon mode,
as schematically depicted in Fig.~\ref{shemaall}d. The Hamiltonian of
this two-impurity Anderson-Holstein model in the particle-hole
symmetric point is
\begin{equation}
\begin{split}
H &= \sum_{k\sigma} \epsilon_k c^\dag_{k\sigma} c_{k\sigma} \\
& + U_1 (n_1-1)^2 + g (a^\dag+a) (n_2-1) + \omega_0 a^\dag a \\
& + \sum_{k\sigma} V^1_k \left(
d^\dag_{1\sigma} c_{k\sigma} + \text{H.c.}
\right) + \sum_{k\sigma} V^2_k \left(
d^\dag_{2\sigma} c_{k\sigma} + \text{H.c.}
\right).
\end{split}
\end{equation}
The presence of the phonon degrees of freedom greatly increases the
degeneracy and thus the number of states that need to be kept in the
truncation step of the NRG \cite{wilson1975}. We used discretization
parameter $\Lambda=3$, the discretization scheme of
Ref.~\onlinecite{campo2005}, and we took into account the
$\sutwo_\mathrm{spin}$ and $\uone_\mathrm{charge}$ symmetries. We
used $\beta=0.75$ \cite{wilson1975, krishna1980a}. We kept up to 5000
states (not taking into account the degeneracies) or the states with
energy below $15T_N$, whichever number is lower. In addition, since
the eigenvalues in NRG are clustered, we took care not to truncate in
the middle of a cluster, so that we do not introduce systematic
errors.  This approach gives converged results which agree for several
values of the parameter $z$ used.

For finite $g$, a pairing moment is induced in the first orbital and a
magnetic moment in the second: we then expect a simultaneous spin and
charge Kondo effect, as in the two-impurity Anderson model.  This is
confirmed by the results for the thermodynamic properties in
Fig.~\ref{fig_f3}. For small $g/\omega_0=0.1$, $U_\mathrm{eff}$ is
small and hence the effective isospin exchange constant $J_C$ is
large: the magnetic exchange constant $J_S$ is then renormalized and
the spin Kondo temperature is reduced, as described in
Sec.~\ref{secIII}.  As $g$ is increased, the charge Kondo temperature
rapidly decreases, while the spin Kondo temperature returns to its
unrenormalized value.  At a particular value of
$g/\omega_0=\sqrt{U/(2\omega_0)}=0.2236$, the spin and charge Kondo
temperatures are the same.

\section{Conclusion and discussion}

We have studied the degree of the spin-charge separation in
single-channel quantum impurity models where the impurity carries both
the spin and isospin degrees of freedom. Using numerical
renormalization group calculations we have confirmed the well known
fact that in the continuum limit, i.e. when the band-width $D$ is much
larger than all other relevant scales in the problem, spin and isospin
degrees of freedom of a single conduction band behave as two decoupled
and independent spin-$1/2$ $\sutwo$ degrees of freedom subject only to
gluing conditions. This implies that the spin and charge Kondo effects
can coexist, the demonstration of which is the main result of our
work. When $D$ is comparable to other scales, lattice effects become
important since a single orbital cannot sustain a spin and isospin
moment at the same time. The separate screening of spin and isospin
degrees of freedom exists even in this limit, but the corresponding
Kondo temperatures are reduced compared to the single-impurity results
due to the renormalization of the Kondo exchange constants.

We have explored the model of a deformable molecule with two active
orbitals, one carrying a local magnetic moment (described by the
Anderson model) and one coupled to a strong phonon mode which induces
local pairing moment (described by the Holstein model). We have shown
that in the generic case the conductance through such a molecule at
the particle-hole symmetric case has non-uniform temperature
dependence: it rises at the higher of the two Kondo temperatures, but
then drops to zero below the lower Kondo temperature. By changing the
gate voltage from its value at the particle-hole symmetric point, the
charge Kondo effect is quenched and the conductance increases back to
the unitary limit if the spin Kondo temperature is much higher than
the charge Kondo temperature, $T_K^S \gg T_K^C$. Conversely, the
magnetic field can be used to quench the spin Kondo effect and the
conductance attains the unitary limit if $T_K^C \gg T_K^S$. Detection
of such behavior could serve as an experimental probe of the
simultaneous spin and charge Kondo effect. The spin-charge Kondo
effect occurs only around the particle-hole symmetric point (i.e. for
precisely tuned gate voltage) and zero magnetic field. In the case of
widely separated charge and spin Kondo temperatures, it would thus
appear as a small region of zero or reduced conductance within a wider
region of the high conductance ``Kondo plateau'' when plotted as a
function of the gate voltage and the magnetic field. In the case of
similar charge and spin Kondo temperatures, it would be difficult to
distinguish (at constant low temperature) the zero conductance due to
simultaneous charge and spin Kondo effects from the Coulomb-blockade
valley. The temperature dependence of the conductance is, however,
different. In the case of Coulomb blockade, conductance monotonously
decreases below the temperature scale of charge excitations. In the
case of simultaneous charge and spin Kondo effects, the decrease due
to Coulomb blockade is followed by increasing conductance that peaks
at the Kondo temperature before it goes to zero.

\begin{acknowledgments}
The authors acknowledge the financial support of the SRA under Grant No.
P1-0044.
\end{acknowledgments}

\bibliography{paper}

\end{document}